\documentclass[multphys,vecphys]{svmult}

\usepackage{makeidx}         
\usepackage{graphicx}        
\usepackage{multicol}        

\makeindex             

\begin{document}

\title*{Dressing of the charge carriers in high-T$_c$ superconductors}
\author{J\"org Fink\inst{1,2}\and
Sergey Borisenko\inst{1}\and
Alexander Kordyuk\inst{1,3}\and
Andreas Koitzsch\inst{1}\and
Jochen Geck\inst{1}\and
Volodymyr Zabolotnyy\inst{1}\and
Martin Knupfer\inst{1}\and
Bernd B\"uchner\inst{1}\and
Helmut Berger\inst{4}}
\authorrunning{J\"org Fink et al.}
\institute{Leibniz Institute for Solid State and Materials
Research Dresden, P.O. Box 270016, 01171 Dresden, Germany
\texttt{J.Fink@ifw-dresden.de} \and Ames Laboratory, Iowa State
University, Ames, Iowa 50011, USA \and Institute of Metal Physics
of the National Academy of Sciences of Ukraine, 03142 Kyiv,
Ukraine \and Institut de Physique de la Mati\`{e}re Complex, Ecole
Politechnique F\'{e}d\'{e}rale de Lausanne, CH-1015 Lausanne,
Switzerland}
%
%
\maketitle

\section{Introduction}
\label{sec:1}
One hundred years ago, in the first of five famous papers
\cite{Einstein} of his {\it annus mirabilis}, Albert Einstein
postulated the dual nature of light, at once particle and wave,
and thereby explained  among other phenomena the photoelectric
effect, originally discovered by H. Hertz \cite{Hertz}. This work
of Einstein was also singled out by the Nobel committee in 1921.
The photoelectric effect has since become the basis of one of the
most important techniques in solid state research. In particular,
angle-resolved photoemission spectroscopy (ARPES), first applied
by Gobeli et al. \cite{Gobeli}, has developed to {\em the}
technique to determine the band structure of solids. During the
last decade, both the energy and the angular resolution of ARPES
has increased by more than one order of magnitude. Thus it is
possible to measure the dispersion very close to the Fermi level,
where the spectral function, which is measured by ARPES, is
renormalized by many-body effects such as electron-phonon,
electron-electron, or electron-spin interactions. The mass
enhancement due to such effects leads to a reduced dispersion and
the finite life-time of the quasi-particles leads to a broadening
of the spectral function. Thus the increase in resolution,
achieved by new analyzers using two-dimensional detectors,
together with new photon sources provided by undulators in
3$^{rd}$ generation synchrotron storage rings and new
cryo-manipulators have opened a new field in ARPES: the
determination of the low-energy many-body properties of solids
which is termed very often the "dressing" of the charge carriers.

In high-T$_c$ superconductors (HTSCs) discovered by Bednorz and
M\"uller \cite{Muller} the many-body effects are supposed to be
particularly strong since these doped cuprates are close to a
Mott-Hubbard insulator or to be more precise to a charge-transfer
insulator \cite {Zaanen}. Since in the normal and the
superconducting state the renormalization effects are  strong, the
HTSCs are a paradigm for the new application of ARPES. Moreover,
since in these compounds the mass enhancement and the
superconducting gap is large, they can be measured using ARPES
even without ultra-high resolution.

On the other hand, the understanding of the renormalization
effects in the HTSCs is vital for the understanding of the
mechanism of high-T$_c$ superconductivity, since the dressing of
the charge carriers may be related with the glue forming the
Cooper pairs. Up to now there is no widely accepted microscopic
theory, although the phenomenon has been discovered already 20
years before. Similar to the conventional superconductors, before
the development of a microscopic theory for the mechanism of
superconductivity, first one has to understand the many-body
effects in the normal state of these highly correlated systems.
ARPES plays a major role in this process. Not only it can
determine the momentum dependent gap. It is at present also the
only method which can determine the momentum dependence of the
renormalization effects due to the interactions of the charge
carriers with other degrees of freedom.

In this contribution we review ARPES results on the dressing of
the charge carriers in HTSCs obtained by our spectroscopy group.
There are previous reviews on ARPES studies of HTSCs \cite{Lynch,
Damascelli, Campuzano}, which compliment what is discussed here.
\section{High-T$_c$ superconductors}
\label{sec:2}
\subsection{Structure and phase diagram}
\label{sec:2.1} It is generally believed that superconductivity is
associated with the two-dimensional CuO$_2$ planes shown in Fig. 1
(a). In these planes Cu is divalent, i.e., Cu has one hole in the
3d shell. The CuO$_2$ planes are separated by block layers formed
by other oxides (see Fig. 1(b)). Without doping, the interacting
CuO$_2$ planes in the crystal form an antiferromagnetic lattice
with a N\'eel temperature of about T$_N$ = 400 K. By substitution
of the ions in the block layers, it is possible to dope the
CuO$_2$ planes, i.e., to add or to remove electrons from the
CuO$_2$ planes. In this review we focus on hole doped systems.
With increasing hole concentration and increasing temperature, the
long-range antiferromagnetism disappears (see the phase diagram in
Fig. 1(c)) but one knows from inelastic neutron scattering that
spin fluctuations still exist at higher dopant concentrations and
higher temperatures.
\begin{figure}
\centering
\includegraphics[height=6cm]{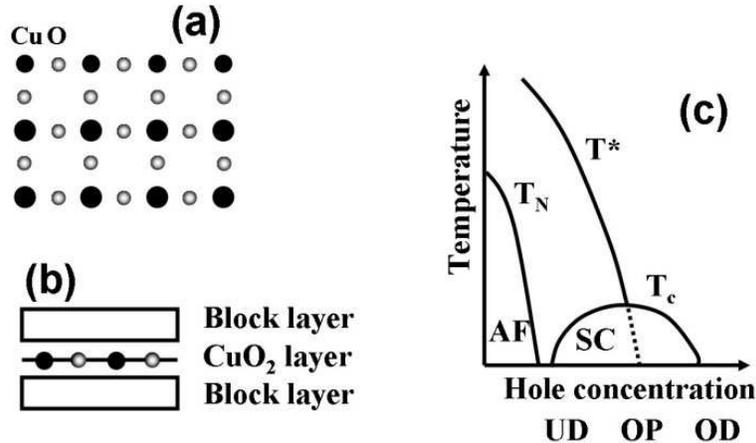}
%
%
\caption{(a) CuO$_2$ plane, (b) CuO$_2$ plane between block
layers, (c) schematic phase diagram of hole-doped cuprates }
\label{fig:1}       
\end{figure}

With increasing dopant concentration the insulating properties
transform into  metallic ones and there is a high-T$_c$
superconducting range. This range is normally divided into an
underdoped (UD), an optimally doped (OP) and an overdoped (OD)
region. Not only the superconducting state but also the normal
state is unconventional. In the UD range there is a pseudogap
between the T$^*$ line and the T$_c$ line. There are various
explanations for the pseudogap \cite{Timusk}: preformed pairs
which have no phase coherence, spin density waves, charge density
waves, or the existence of a hidden order, caused, e.g. by
circulating currents \cite {Varma2}. At low temperatures in the OP
region the T$^*$ line is very often related to a quantum critical
point near the OP region. Possibly related to this quantum
critical point, in the OP range, the normal state shows rather
strange properties such as a linear temperature dependence of the
resistivity over a very large temperature range or a temperature
dependent Hall effect. Only in the OD range the system behaves
like a normal correlated metal showing for example a quadratic
temperature dependence of the resistivity.

\subsection{Electronic Structure}
\label{sec:2.2}

In the following we give a short introduction into the electronic
structure of cuprates. We start with a simple tight-binding
bandstructure of a CuO$_2$ plane using for the beginning three
hopping integrals, one between 2 neighboring Cu sites along the
Cu-O bonding direction ($t$), one for a hopping to the second
nearest Cu neighbor along the diagonal ($t'$), and one for the
hopping to the third nearest neighbor ($t''$). The corresponding
bandstructure is given by
\begin{eqnarray}
E(\vec{k}) &=&\Delta \epsilon  -2t[cos(k_xa)+cos(k_ya)]+4t'cos(k_xa)cos(k_ya)\nonumber\\
& &-2t''[cos(2k_xa)+cos(2k_ya)]
\end{eqnarray}
where $a$ is the length of the unit cell and $\Delta \epsilon$
fixes the Fermi level. This  two-dimensional bandstructure is
displayed in Fig. 2 (a) for $t'/t=-0.3$ , a value which is
obtained from bandstructure calculations \cite {Andersen}, and
both $t''$ and $\Delta \epsilon$ equal to zero . It has a minimum
in the center ($\Gamma$ ) and a maxima at the corners of the
Brillouin zone (e.g. at $(k_x,k_y)=(\pi,\pi)/a\equiv(\pi,\pi)$).
Furthermore there are saddle points, e.g. at ($k_x,k_y)=(\pi,0)$.
In the undoped system there is one hole per Cu site and therefore
this band should be half filled. This leads to a Fermi level just
above the saddle points (see Fig. 2(a)). The Fermi surface
consists of rounded squares  around the corners of the Brillouin
zone (see Fig. 2(b)). Upon hole doping the Fermi level moves
towards the saddle point. It is interesting that for vanishing
$t'$ the Fermi surface would be quadratic and there would be no
parallel sections (which could lead to a nesting) along $x$ or $y$
but along the diagonal. There are two special points on the Fermi
surface (see Fig. 2 (b)), which are also at the focus of most of
the ARPES studies on HTSCs. There is the nodal point at the
diagonal (N in Fig. 2(b)), where the superconducting order
parameter is zero and the antinodal point (where the
($\pi,0)$-($\pi,\pi)$ line cuts the Fermi surface), where the
superconducting order parameter is believed to reach a maximum (AN
in Fig. 2(b) \cite {Ding, Borisenko2}.
\begin{figure}
\centering
\includegraphics[height=8cm]{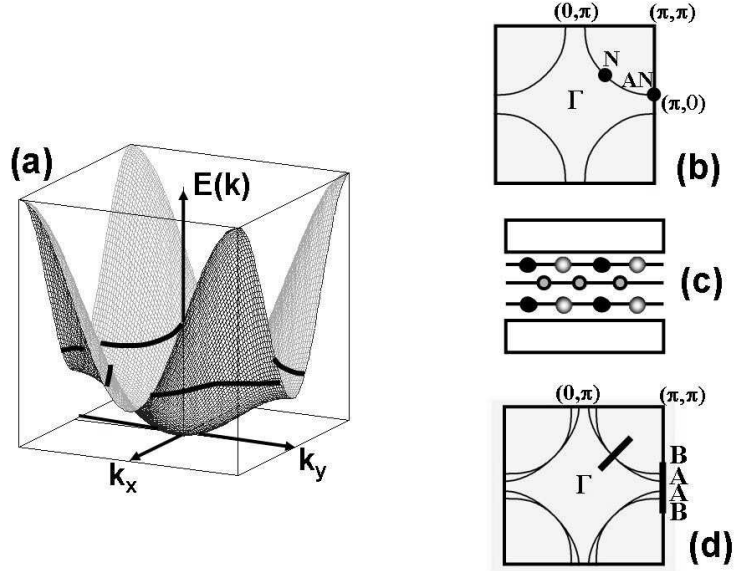}
\caption{(a) Tight-binding bandstructure of the CuO$_2$ plane. (b)
Fermi surface of a CuO$_2$ plane. N: nodal point, AN: antinodal
point, (c) Bilayer system between block layers composed of two
CuO$_2$ planes separated by one ionic layer, (d) Fermi surfaces of
a bilayer system, B(A): (anti)bonding band. Thick solid lines: k
values along which most of the present ARPES studies have been
performed.}
\label{fig:2}       
\end{figure}

In many cuprates there is not just a single but several CuO$_2$ planes between the block
layers. In these systems the CuO$_2$ planes are separated by  additional ionic layers. This is illustrated
for a bilayer system in Fig. 2(c). Such a bilayer system is
for example Bi$_2$Sr$_2$CaCu$_2$O$_8$  which is the
Drosophila for ARPES studies of HTSCs. In this compound the block layers are
composed of BiO and SrO planes, while the ionic layer separating
the two CuO$_2$ planes  consist of  Ca\raisebox{1ex}{2+} layers. Doping is achieved
in this compound by additional O atoms in the block layers.
In those bilayer systems there is an interaction between the two adjacent  CuO$_2$ planes
which leads to a finite hopping integral t$_\perp$. This causes  an additional
term in the tight-binding calculations
\begin{equation}
E(\vec{k})_{\perp} = \pm t_{\perp}[cos(k_xa)-cos(k_ya)]^2/4
\end{equation}
leading to a splitting into a bonding and an antibonding band.
This splitting is small at the nodal point \cite {Kordyuk6} and it
is largest at the antinodal point. In Fig. 2(d) we have
illustrated this splitting of the Fermi surface caused by the
interaction of the two CuO$_2$ planes.

The independent particle picture, describing just the interactions
with the ion lattice and the potential of a homogeneous conduction
electron distribution, is of minor use for the undoped systems
since we know that those are not metallic but insulating.  This
comes from the Coulomb interaction U of two holes on the same Cu
site which prohibits hopping of holes from one Cu site to the
other. It causes the insulating behavior of undoped and slightly
doped cuprates. The large on-site Coulomb repulsion of two holes
on a Cu site is also responsible for the fact that the additional
holes produced upon doping are formed on O sites \cite {Nucker1}.
The 2 eV energy gap is then a charge-transfer gap \cite {Zaanen}
between O 2p and Cu 3d states. Only when more and more holes are
introduced into the CuO$_2$ planes is hopping of the holes
possible and correlation effects get less important.
\section{Angle-resolved photoemission spectroscopy}
\label{sec:3}
\subsection{Principle}
\label{sec:3.1} In photoemission spectroscopy monochromatic light
with an energy $h\nu $ is shined onto a surface of a solid and the
intensity as well as the kinetic energy, E$_{kin}$, of the
outgoing photoelectrons is measured. Using the explanation of the
photoelectric effect \cite{Einstein} one can obtain the binding
energy of the electrons in the solid:
\begin{equation}
E_B=h\nu-\Phi -E_{kin}\equiv -E.
\end{equation}
Here $\Phi $ is the workfunction. The charge carriers in HTSCs
show a quasi-two-dimensional behavior. When the surface is
parallel to the CuO$_2$ planes, the momentum
$\hbar$k$_{\parallel}$ of the photoelectron is conserved when
passing through the surface and thus this momentum is determined
by the projection of the total momentum of the photoelectron to
the surface:
\begin{equation}
\hbar k_{\parallel}= \sqrt{ 2mE_{kin}}\sin\theta .
\end{equation}
Here $\theta$  is the angle between the direction of the photoelectron
 in the vacuum and the surface normal.

There are numerous treatises of the photoelectron process in the
literature where the limitations of the models which describe it
are discussed \cite{Hufner}. They are not repeated in this
contribution. Rather the essential points for the analysis of
ARPES studies on the dressing of the charge carriers in HTSCs are
restated. It is assumed that the energy and momentum dependence of
the photocurrent in ARPES studies can be described by
\begin{equation}
I(E ,k) \propto M^2A(E ,k)f(E) +B(E ,k)
\end{equation}
where $M=<\psi _f|H'|\psi _i>$ is a matrix element between the initial and the
final state and $H'$ is a dipole operator. $A(E ,k)$ is the spectral function
which is the essential result in ARPES studies.
$ f(E) =1/[exp(E /k_BT)+1]$ is the Fermi function which takes into account
that only occupied states are measured and $B(E ,k)$
is an extrinsic background coming from secondary electrons.  For
a comparison of calculated data with experimental data, the former
have to be convoluted with the energy and momentum resolution.

The dynamics of an electron in an interacting system can be described by a
Green's function \cite{Mahan}
\begin{equation}
G(E ,k)=\frac{1}{E -\epsilon _k-\Sigma(E ,k)}.
\end{equation}
$\Sigma(E,k)$=$\Sigma'(E,k)$+i$\Sigma''(E,k)$ is the complex
self-energy function which contains the information on the
dressing, i.e., on what goes beyond the independent-particle
model. $\epsilon _k$ gives the dispersion of the bare particles
without many-body interactions. The spectral function can be
expressed \cite{Hedin,Almbladh} by
\begin{equation}
A(E ,k)=-\frac{1}{\pi}ImG(E ,k)=-\frac{1}{\pi}\frac{\Sigma''(E ,k)}
{[E -\epsilon _k-\Sigma'(E ,k)]^2+[\Sigma''(E ,k)]^2}
\end{equation}
For  $\Sigma=0$, i. e., for the non-interacting case, the Greens
function and thus the spectral function is a delta-function at the
bare-particle energy $\epsilon _k$. Taking interactions into
account, the spectral function given in Eq. (7) is a rather
complicated function. On the other hand, in many cases only local
interactions are important which leads to a $k$-independent or
weakly $k$-dependent self-energy function. Furthermore, in the
case of not too strong interactions, often  quasi-particles with
properties still very close to the bare particles, can be
projected out from the spectral function. To perform this
extraction one expands the complex self-energy function around the
bare particle energy $\epsilon _k$:
$\Sigma(E)\approx\Sigma(\epsilon _k)+\partial \Sigma(E)/
\partial E\big|_{E=\epsilon _k}(E-\epsilon _k)$.
Very often one introduces the coupling constant $\lambda $ =
-$\partial \Sigma''(E)/\partial E|_{E=\epsilon _k}$ and the
renormalization constant $Z=1+\lambda $. Note that in many
contributions in the literature Z$^{-1}$ is replaced by Z.
Neglecting  the partial derivative of $\Sigma''(E)$  one obtains
for the spectral function
\begin{equation}
A(E,k)_{coh}=-\frac{1}{\pi}Z(\epsilon _k)^{-1}\frac{Z(\epsilon _k)^{-1}\Sigma''(\epsilon _k)}
{[E -\epsilon _k-Z(\epsilon _k)^{-1}\Sigma'(\epsilon _k)]^2+
[Z(\epsilon _k)^{-1}\Sigma''(\epsilon _k)]^2}
\end{equation}
This is the coherent fraction of the spectral function and its
spectral weight is given by $Z^{-1}$. It is called coherent
because it describes a (quasi-)particle which is very similar to
the bare particle. Instead of a delta-function we have now a
Lorentzian. The energy of the quasi-particle  is determined by the
new maximum of the spectral function which occurs at $E =\epsilon
_k-Z^{-1}\Sigma'(\epsilon _k)$. The life-time of the
quasi-particle is determined in a cut at constant k by the FWHM of
the Lorentzian which is given by $\Gamma =
2Z^{-1}\Sigma''(\epsilon _k)$.

Close to the Fermi we can assume that the real part of the
self-energy is linear in energy, i.e., $\Sigma'(\epsilon _k)=
-\lambda\epsilon _k$. For the renormalized energy of the
quasi-particle, we now obtain $E_k =\epsilon _k/(1+\lambda)$. Thus
close to the Fermi level we have in the case of a linear real part
of the self-energy a renormalization by a factor of $1+\lambda$ or
in other words, due to  the interactions we have for the coherent
quasiparticles a mass enhancement $m^*=(1+\lambda)m$.

The incoherent part of the spectral function, the spectral weight
of which is given by $1-Z^{-1}$, contains all the spectral weight
which cannot be described by the Lorentzian close to the bare
particle energy, e.g., satellites. $Z^{-1}$ also determines the
size of the jump at $k_F$ of the momentum distribution
 $n(k)$, which can be calculated from the energy integral of the
 spectral function $A(E,k)$. Thus if the jump in $n(k)$
comes to zero, at this very point the quasiparticles weight
$Z^{-1}$  vanishes logarithmically as one approaches the Fermi
level. For such an electron liquid the term ``marginal'' Fermi
liquid \cite {Varma} has been introduced. This is related to
another condition for the existence of (coherent) quasiparticles
\cite{Pines, Imada}. The finite lifetime implies an uncertainty in
energy. Only if this uncertainty is much smaller than the binding
energy ($\Sigma ''/E\rightarrow0$) the particles can propagate
coherently and the concept of quasiparticles has a physical
meaning.

In principle, performing constant-k scans, commonly called energy
distribution curves (EDCs), one can extract the spectral function
along the energy axis and using Eq. (7) one can derive the complex
self-energy function. In reality there is a background, the exact
energy dependence of which is not known. In addition, close to the
Fermi energy there is the energy-dependent Fermi function. These
problems are strongly reduced when performing constant-energy
scans, usually called momentum distribution curves (MDCs)
\cite{Valla}. Close to the Fermi level the bare particle
bandstructure can be expanded as $\epsilon _k= v_F\hbar(k-k_F)$.
Assuming again a weakly k-dependent $\Sigma (E,k)$, the spectral
function along the particular k-direction is a Lorentzian (see Eq.
(7)). The width is given by $\Sigma ''/v_F\hbar$ and from the
shift relative to the bare particle dispersion one can obtain the
real part $\Sigma'$. This evaluation is much less dependent on a
weakly k-dependent background and on the Fermi function.

\subsection{Spectral function in the normal state}
\label{sec:3.2}
In a real solid there are several contributions to the self-energy. The important ones, related to
inelastic scattering processes can be reduced  to
contributions which are related to bosonic excitations (see Fig. 3).
\begin{figure}
\centering
\includegraphics[height=4cm]{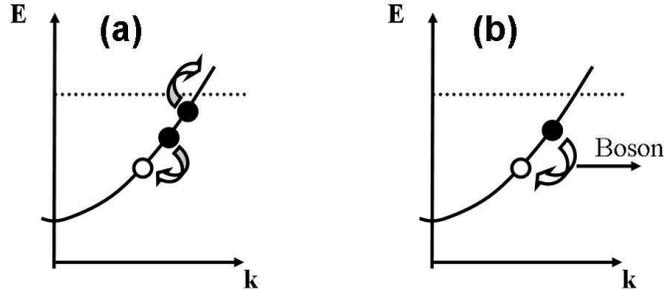}
\caption{Bosonic excitations contributing to the finite lifetime
of a photohole in metallic solids. (a) electron-hole excitations,
(b) discrete bosonic mode. The dashed line corresponds to the
Fermi level}
\label{fig:3}       
\end{figure}
In the case where the boson is a particle-hole excitation, which
is depicted in Fig.~3(a), a photoelectron hole is filled by a
transition from a higher energy level and the energy is used to
excite an Auger electron above the Fermi level. The final state is
thus a photoelectron hole scattered into a higher state plus an
electron-hole pair. For a normal Fermi liquid of a
three-dimensional solid at T = 0 phase space arguments and the
Pauli principle lead to the complex self-energy function
$\Sigma=\alpha E-i\beta E^2$. In a two-dimensional solid the
imaginary part of $\Sigma$ changes from a quadratic energy
dependence to $\beta'E^2ln| E/E_F|$ \cite{Hodges} which is similar
to the 3D case only as long as $E$ is much smaller than the
bandwidth. Increasing the interactions more and more, associated
with a reduction of $Z^{-1}$, changes the self-energy function.
 For $Z^{-1}=0$ where the the spectral
weight of the quasi-particles disappears one reaches the above
mentioned marginal Fermi liquid \cite{Varma}. In this case the
self-energy is given by $\Sigma=\lambda_{MFL}[E
ln|x/E_c|+i(\pi/2)x] $ where $x=max(E,k_BT)$ and $E_c$ is a cutoff
energy taking into account the finite width of the conduction
band. This self-energy function is a phenomenological explanation,
among others, of the linear temperature dependence of the
resistivity observed in optimally doped HTSCs, since the imaginary
part of the self-energy and thus the inverse scattering rate is
linear in $T$.

Besides the particle-hole excitations described above, the
photohole may be scattered to higher (lower) energies by the
emission (absorption) of a discrete boson. This is illustrated in
Fig. 3(b) for the emission of a bosonic excitation. Such discrete
bosonic excitations may be phonons, spin excitations, plasmons,
excitons etc. The relevant excitations are listed in Table 1
together with their characteristic energies in optimally doped
HTSCs.

\begin{table}
\centering
\caption{Bosonic excitations which couple to the charge carriers together with
their characteristic energies in HTSCs}
\label{tab:1}       
%
%
\begin{tabular}{llr}
\hline\noalign{\smallskip}
system &\hspace{.5cm} excitations &\hspace{1cm}characteristic energy(meV)  \\
\noalign{\smallskip}\hline\noalign{\smallskip}
ion lattice &\hspace{.5cm} phonons & 90\hspace{2cm} \\
spin lattice/liquid &\hspace{.5cm} magnons & 180\hspace{2cm} \\
e-liquid &\hspace{.5cm} plasmons & 1000\hspace{2cm} \\
\noalign{\smallskip}\hline
\end{tabular}
\end{table}

\begin{figure}
\centering
\includegraphics[height=8cm]{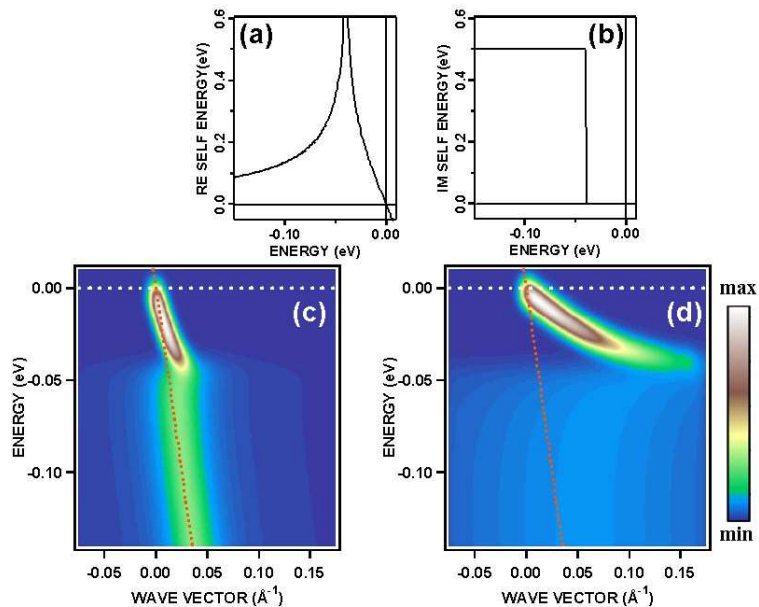}
\caption{Real part (a) and imaginary part (b) of the self-energy
function for a coupling to a mode $\Omega_0$ = 40 meV and a
coupling constant $\lambda $=8. Spectral function $A(E,k)$ for
$\lambda =1$ (c) and  $\lambda =8$ (d) in the normal state.}
\label{fig:4}       
\end{figure}

The self-energy function for a coupling of the charge carriers to
a bosonic mode for the case that the energy of the mode is much
smaller than the band width has been treated by Engelsberg and
Schrieffer \cite{Schrieffer}. The assumption of a strong screening
of the bosonic excitations is probably adequate for the doped
HTSCs but probably not for the undoped or slightly doped parent
compounds \cite{KShen, Roesch}. In the well-screened case,
$\Sigma''$ is zero up to the mode energy $\Omega_0$. This is
immediately clear from Fig. 3 (b) since the photohole can only be
filled when the binding energy is larger than $\Omega_0$.
$\Sigma''$ is constant above the mode energy (see Fig. 4(b)).
Performing the Kramers-Kronig transformation one obtains
$\Sigma'$, which is given by $\Sigma'= (\lambda \Omega
_0/2)ln|(E+\Omega _0)/(E-\Omega _0)|$ (see Fig. 4 (a)). It shows a
logarithmic singularity at the mode energy, $\Omega_0$. At low
energies there is a linear energy dependence of $\Sigma'$ and the
slope determines the coupling constant $\lambda$. In this model it
is related to the imaginary part of the self-energy function
$\Sigma ''(|E|>\Omega _0) \equiv \Sigma''(-\infty)$ , by $\lambda
= -\Sigma ''(-\infty)/(\pi \Omega _0/2)$. From this it is clear
that for a given $\Omega_0$ both $\lambda $ and $\Sigma
''(-\infty)$ are a measure of the coupling strength to the bosonic
mode.

In Fig. 4(c) and (d) we have displayed the calculated spectral
function for $\lambda =1$ and $\lambda =8$, respectively. Compared
to the bare particle dispersion, given by the red dashed line, for
$|E| < \Omega_0$ there is a mass renormalization, i.e., a reduced
dispersion and no broadening, except the energy and momentum
resolution broadening, which was taken to be 5 meV and 0.005
\AA$^{-1}$, respectively . For $|E| > \Omega _0$, there is a
back-dispersion to the bare particle energy. Moreover, there is a
broadening due to a finite $\Sigma''$, increasing with increasing
$\lambda$. For large $\lambda$, the width for constant E scans is,
at least up to some energy,  larger than the energy of the charge
carriers and therefore they can be called incoherent (see Sect.
\ref{sec:3.2}) in contrast to the energy range $|E| < \Omega _0$
or at very high binding energies, where the width is much smaller
than the binding energy and where they are coherent \cite
{Schrieffer}. The change in the dispersion is very often termed a
"kink" but looking closer at the spectral function, in particular
for high $\lambda$, it is a branching of 2 dispersion arms.

\subsection{Spectral function of solids in the superconducting state}
\label{sec:3.3}

For the description of the spectral function in the
superconducting state, two excitations have to be taken into
account: the electron-hole and the pair excitations. This leads to
a (2x2) Green's function \cite {Nambu}. Usually the complex
renormalization function
\begin{equation}
Z(E,k)=1-\Sigma (E,k)/E,
\end{equation}
is introduced. For the one-mode model, the self-energy of the
superconducting state corresponds to the self-energy of the normal
state in which $\Omega_0$ is replaced by $\Omega_0+\Delta $. This
can be easily seen from Fig. 3(b) and assuming a gap opening with
the energy $\Delta $. The coupling constant in the superconducting
state, $\lambda_{sc}$, is related to the renormalization function
by $\lambda_{sc}=Z(0)-1$. For the Auger process shown in Fig. 3(a)
the onset of the scattering rate is at 3$\Delta $. The reason for
this is that the bosonic (e-h) excitations have in this case a
lower limit of 2$\Delta $. The complex spectral function is given
by \cite{Scalapino}
\begin{equation}
A(E,k)=-\frac{1}{\pi}Im\frac{Z(E,k)E+\epsilon_
k}{Z(E,k)^2(E^2-\Delta (E,k)^2)-\epsilon_k^2}.
\end{equation}
In general, $\Delta (E,k)$ is also a complex function.  In Fig. 5 we show for the one-mode model the calculated
spectral function in the superconducting state using the same energy and momentum resolutions and the
same mode energy as before. The imaginary part of $\Delta$ was neglected and the real part was set
to 30 meV. One clearly realizes the BCS-Bogoliubov-like back-dispersion at
the gap energy $\Delta$ and besides this, a
total shift of the dispersive arms by the gap energy. Thus the branching energy occurs
at  $\Omega_0+\Delta $.
\begin{figure}
\centering
\includegraphics[height=8cm]{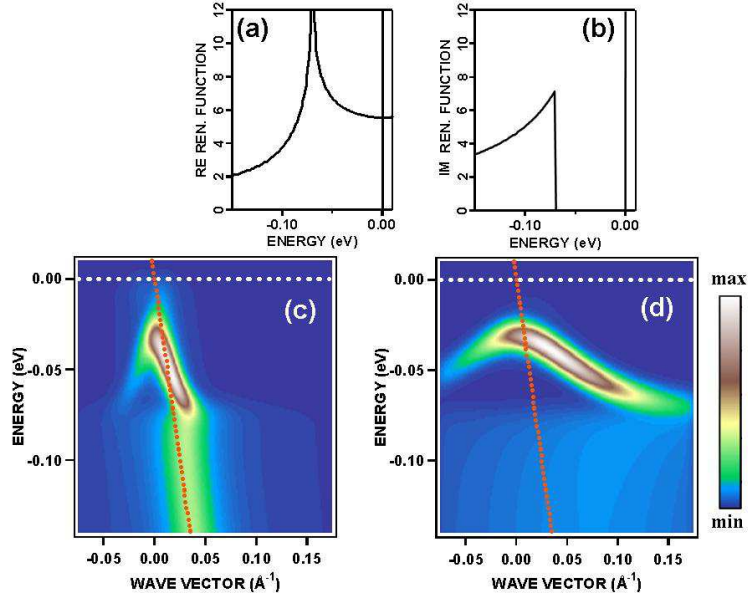}
\caption{The real part (a) and imaginary part (b) of the
renormalization function Z(E) for $\Omega _0$=40meV, $\Delta $=30
meV, and $\lambda _{sc}$ = 5. Spectral function $A(E,k)$ for a
coupling constant $\lambda _{sc} =1$ (c) and  $\lambda _{sc}=5$
(d) in the superconducting state.}
\label{fig:5}       
\end{figure}

Looking at the phase diagram in Fig. 1(c) it is clear that the
HTSCs are very close to a transition into a Mott-insulating state
and therefore we expect a large fraction of incoherent spectral
weight in the normal state. This, however changes when going into
the superconducting state where for $min(\Delta+\Omega
_0,3\Delta)>|E|>\Delta $ the incoherent states are transformed
into coherent ones. The reason for this is that in the
superconducting state a gap opens for $\Sigma ''$ for $|E|<
3\Delta$ (e-h scattering rate) or for $|E|<\Omega_0+\Delta$
(bosonic scattering rate).

The dispersion is given by \cite {Chubukov}:
\begin{equation}
[ReZ(E,k)]^2[E^2-\Delta (E,k)^2]-\epsilon_k^2 = 0.
\end{equation}
In the conventional superconductors the mode energy is much larger
than the gap and therefore for $|E|$ slightly larger than $\Delta$
, Z, and thus $\lambda_{sc} $, is constant. In this case Eq. (11)
yields for the maxima of the spectral function
\begin{equation}
E=\sqrt{\Delta ^2+\epsilon _k^2/(1+\lambda_{sc})^2}.
\end{equation}
For HTSC the gap is comparable to the mode energy and therefore
Eq. (12) is no longer valid and the full Eq. (11) should be used
to fit the dispersion. Then $\lambda_{sc}$  is related to the
normal state $\lambda_n$ (from $\lambda_n = (Z(0)-1)|_{\Delta
=0})$ by $\lambda_n = \lambda_{sc}(\Omega_0+\Delta )/\Omega_0$. It
is this $\lambda_n$ which should be considered when comparing the
coupling strength of the charge carriers to a bosonic mode of
HTSCs and conventional superconductors.

When one measures an EDC at $k_F$ a peak is observed followed by a
dip and a hump. Such an energy distribution is well known from
tunnelling spectroscopy in conventional superconductors which was
explained in terms of a coupling of the electrons to phonons. A
closer inspection indicates for the one-mode-model that at $k_F$
the peak is followed by a region of low spectral weight and a
threshold, which appears at $\Omega_0+ \Delta$. Far away from
$k_F$ this threshold is not contaminated by the tails of the peak.

\subsection{Experimental}
\label{sec:3.4} During the last decade ARPES has experienced an
explosive period of qualitative and quantitative improvements.
Previously ARPES was performed by rotating the analyzer step by
step. In this way an enormous amount of information was lost
because only one angle of the emitted photoelectrons was recorded.
The development of the so-called ``angle mode" \cite{Martensson},
applied in the new generation of SCIENTA analyzers, allows the
simultaneous recording of both an energy and an angle range. This
was achieved by a multielement electrostatic lens system, by which
each photoemission angle was imaged to a different spot of the
entrance slit of the a hemispherical, electrostatic deflection
analyzer. This angular information is then transferred to the exit
of the analyzer and the energy and angle dispersion is recorded by
a two-dimensional detector consisting of a microchannel plate, a
phosphor plate, and a charge coupled device detector. This caused
an improvement both of the energy and the momentum resolution by
more than one order of magnitude and  an enormous improvement of
the detection efficiency, leading to a very strong reduction of
measuring time. But not only new analyzers and detectors lead to a
huge progress of the ARPES technique. Also new photon sources such
as undulators in synchrotron storage rings \cite{Koch}, new
microwave driven He discharge lamps, and new cryo-manipulators
contributed to the rapid development of the method.

The measurements presented in this contribution were performed
with SCIENTA SES 200 and 100 analyzers using the above-mentioned
angle mode. The photon sources used were a high-intensity He
resonance GAMMADATA VUV 5000 lamp or various beamlines, delivering
linearly or circularly polarized light in a wide energy range
between 15 and 100 eV: the U125/1 PGM beamline at BESSY
\cite{Follath}, the 4.2R beamline "Circular Polarization" at
ELETTRA, or the beamline SIS at  the SLS. The angular rotation of
the sample was achieved by a purpose built high-precision
cryo-manipulator which allows the sample to be cooled to 25 K and
a computer-controlled angular scanning around three perpendicular
axes in a wide range of angles with a precision of 0.1$^{\circ}$.
The energy and the angle/momentum resolutions were set in most
cases in the ranges 8-25 meV and 0.2$^{\circ}$ /0.01-0.02
\AA$^{-1}$, respectively, which is a compromise between energy and
momentum resolution and intensity.

Almost all results presented in this review were obtained from
high-quality and well characterized single crystals of
(Bi,Pb)$_2$Sr$_2$CaCu$_2$O$_{8+\delta} $ (Bi2212). The reason for
this is the following. There is a van der Waals bonding between
two adjacent BiO planes and therefore it is easy to cleave the
crystals. Upon cleaving, no ionic or covalent bonds are broken
which would lead to polar surfaces and to a redistribution of
charges at the surface. Moreover, we know from bandstructure
calculations that among all HTSCs, the Bi-compounds have the
lowest k$_z$ dispersion, i.e., they are very close to a
two-dimensional electronic system. This is very important for the
evaluation of the ARPES data. Probably on all other HTSCs, upon
cleaving there is a redistribution of charges and possibly a
suppressed superconductivity at the surface. The bilayer system of
the Bi-HTSC family is complicated by the existence of 2 bands at
the Fermi surface. On the other hand, it is that system where the
whole superconducting range from the UD to the OD range can be
studied. The system without Pb has a further complication. It has
a superstructure along the b-axis leading in ARPES to diffraction
replicas which complicate the evaluation of the data
\cite{Borisenko1,Legner}. In order to avoid this, about 20 \% of
the Bi ions were replaced by Pb which leads to superstructure-free
samples.
\begin{figure}
\centering
\includegraphics[height=5cm]{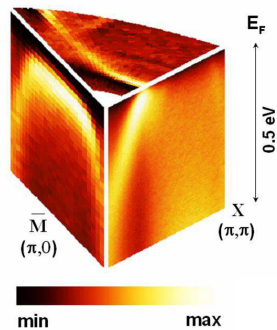}
\caption{Photoelectron intensity of a
(Bi,Pb)$_2$Sr$_2$CaCu$_2$O$_{8+\delta }$ single crystal in the
three-dimensional $(E, k_x, k_y)$ space measured at room
temperature by ARPES. }
\label{fig:6}       
\end{figure}

The potential of the new generation ARPES technique is illustrated
in Fig. 6 where we show  room temperature data of OP  Bi2212 in
the three-dimensional $(E, k_x, k_y)$ space.  The fourth dimension
is symbolized by the color scale, representing the photoelectron
intensity. The right front plane of the section shown in Fig. 6
was taken simultaneously by setting the k-vector parallel to the
$\Gamma -(\pi, \pi)$ direction. Then the sample was turned step by
step until the k-vector was parallel to the $\Gamma -(\pi, 0)$
direction thus sampling 100000 data points of the whole section.

Such a ``piece of cake" can be cut along different directions. A
horizontal cut at the Fermi level yields the Fermi surface. A
vertical cut along a certain k-direction yields the
``bandstructure" (the bare particle dispersion plus the
renormalization) along this direction. In these data, the
essential points of the bandstructure shown in Fig. 2 are
reproduced. Along the $\Gamma-(\pi, \pi )$ direction there is a
crossing of the Fermi level at the nodal point (close to $(\pi/2,
\pi/2 )$). Along the $\Gamma-(\pi, 0 )$ direction there is no
crossing of the Fermi level but the saddle point is realized just
below E$_F$.
\section{The bare-particle dispersion}
\label{sec:4} In order to extract the dressing of the charge
carriers due to the many-body effects from the ARPES data, one has
to know the bare-particle dispersion, i.e., the dispersion which
is only determined by the interaction with the  ions and the
potential due to a homogeneous conduction electron distribution.
We have suggested three different ways to obtain the bare-particle
band structure.

The first one starts with the Fermi surface measured by ARPES. How
to measure those has been already described in Sect.
\ref{sec:3.4}. In Fig. 7 we show ARPES measurements of the Fermi
surface of Bi2212 for various dopant concentrations \cite
{Kordyuk1}. Using a commonly employed empirical relation \cite
{Tallon} between T$_c$ and the hole concentration, $x$, determined
from chemical analysis, the measured samples cover a doping range
of $x$=0.12 to 0.22. The measurements were performed at room
temperature.
\begin{figure}
\centering
\includegraphics[height=5cm]{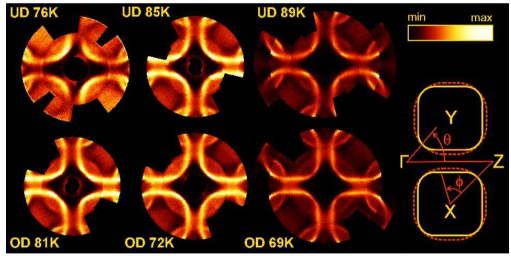}
\caption{Fermi surfaces of (Bi,Pb)$_2$Sr$_2$CaCu$_2$O$_{8+\delta} $ having
various dopant concentrations and T$_c$ (indicated in the panels in units of K) measured
by ARPES at room temperature. Upper row: underdoped (UD) samples, lower row: overdoped (OD) samples.}
 \label{fig:7}       
\end{figure}

Before we come to the evaluation of the bare-particle dispersion,
we make some remarks on the character of the measured Fermi
surfaces. Firstly, the topology does not change, which means that
within the studied doping range there is no transition from a
hole-like to an electron-like surface. Secondly, the shape of the
Fermi surface around $(\pi ,\pi )$ changes from being quite
rounded at low doping to taking on the form of a square with well
rounded corners at higher doping. This is exactly what is expected
within a rigid band approximation and looking at Fig. 2. At low
doping we are far away from the saddle point and we expect a more
rounded Fermi surface. At higher doping we move E$_F$ closer to
the saddle point leading to a more quadratic Fermi surface.
Thirdly, in underdoped samples, there is an intensity reduction
close to $(\pi,0)$ although the intensities are normalized to the
total intensity along the particular k-direction to reduce effects
due to the k-dependence of the matrix element in Eq. (5). This
reduction in spectral weight is related to the formation of the
pseudo-gap below T$^*$, which is above room temperature in the
underdoped samples. This can be treated as a formation of arcs
around the nodal points for low dopant concentrations.

It is possible to fit the measured Fermi surface using Eq. (1).
Such a fit is shown for an optimally doped sample on the right
hand side of Fig. 7 by a yellow line. Only recently, due to the
improved resolution, the bilayer splitting in HTSCs has been
resolved \cite{Feng,Chuang}, while in low-resolution data the
non-detection of this splitting was ascribed to a strong
incoherence of the electronic states close to $(\pi ,0)$. From
calculations of the energy dependence of the matrix element in Eq.
(5) \cite {Bansil, Fujimori} and from systematic
photon-energy-dependent measurements (see below) we know that for
the photon energy $h\nu $= 21.2 eV the matrix element for the
bonding band is more than a factor 2 larger than for the
antibonding band. Therefore, we see in Fig. 7 mainly the Fermi
surface of the bonding band. Utilizing other photon energies, the
bilayer splitting can be clearly resolved, even for UD samples
\cite {Kordyuk1}. The red rounded squares in Fig. 7 illustrates
the Fermi surface of the antibonding band. From the evaluation of
the area of the Fermi surface and taking into account the bilayer
splitting it is possible to derive the hole concentration which
nicely agrees with those values derived from T$_c$ using the
universal relation, mentioned above. This is an important result
supporting the validity of Luttinger's theorem (the volume of the
Fermi surface should be conserved upon switching on the
interactions) within the studied concentration range. Finally, we
mention the existence of a shadow Fermi surface which corresponds
to a ($\pi ,\pi $) shifted (normal) Fermi surface in the Fermi
surface data, shown in Fig. 7. After its first observation
\cite{Aebi}, it was believed to occur due to the emission of spin
fluctuations. More recent measurements indicate that its origin is
related to structural effects \cite{Koitzsch1}.

Now we come back to the determination of the bare-particle band
structure. Assuming that the self-energy effects at E$_F$ are
negligible (which is supported by the experimental result that the
Luttinger theorem is not violated in the concentration range under
consideration), it is possible to obtain information on the
unrenormalized bandstructure from the Fermi surface. By fitting
the Fermi surface with a tight-binding bandstructure, one obtains
relative values of the hopping integrals, i.e., the hopping
integrals $t'$, $t''$, and $t_{\perp}$ normalized to $t$. To
obtain the absolute values we have measured the spectral function
along the nodal direction. From the measured widths at constant
energies one can derive the imaginary part of the self-energy
function. Performing a Kramers-Kronig transformation, it is
possible to derive the real part of $\Sigma$ and using Eq. (7) it
is possible to calculate the bare-particle dispersion from
$\epsilon _k=E_M- \Sigma'$ where E$_M$ is the measured dispersion
(see Sect. \ref{sec:5} ). In this way \cite{Kordyuk2} the absolute
values of the hopping integrals for an UD and an OD sample has
been obtained (see Table 2).
\begin{table}
\centering
\caption{Tight-binding parameters for an underdoped and overdoped (Bi,Pb)$_2$Sr$_2$CaCu$_2$O$_{8+\delta} $
sample }
\label{tab:2}       
%
%
\begin{tabular}{lccccc}
\hline\noalign{\smallskip}
sample &\hspace{.5cm} $t$(eV) &\hspace{.2cm}$t'$(eV)&\hspace{.2cm}$t"$(eV)&\hspace{.2cm}$t_\perp$(eV)&\hspace{.2cm}$\Delta \epsilon$   \\
\noalign{\smallskip}\hline\noalign{\smallskip}
UD 77 K &\hspace{.5cm} 0.39 &\hspace{.2cm} 0.078&\hspace{.2cm}0.039&\hspace{.2cm}0.082&\hspace{.2cm}0.29 \\
OD 69 K &\hspace{.5cm} 0.40 &\hspace{.2cm} 0.090&\hspace{.2cm}0.045&\hspace{.2cm}0.082&\hspace{.2cm}0.43 \\
\noalign{\smallskip}\hline
\end{tabular}
\end{table}

A second way to determine the bare-particle bandstructure is to evaluate the anisotropic plasmon
dispersion which was measured by electron energy-loss spectroscopy
for momentum transfers parallel to the CuO$_2$ planes
 \cite{Nucker,Grigoryan}. This plasmon
dispersion is determined by the projection of the Fermi velocity on the plasmon propagation
directions, which could be varied in the experiment.
Since the (unscreened) plasmon energy is at about 2 eV, these excitations are considerably higher than the
renormalization energies (see Table 1) and therefore the plasmon dispersion is determined by the
unrenormalized, averaged Fermi velocity. It is thus possible to fit the momentum dependence of the averaged
Fermi velocity by a tight-binding bandstructure. Similar
hopping integrals as those shown in Table 2 were obtained for an optimally doped
sample. Of course no information on the bilayer splitting could be obtained from those measurements.

Finally, a third way to obtain the bare-particle bandstructure is
to look at the LDA bandstructure calculations \cite{Andersen}. It
is remarkable that the tight-binding  parameters, obtained from a
tight-binding fit of the LDA bandstructure, are very similar to
those given in Table 2.
\section{The dressing of the charge carriers at the nodal point}
\label{sec:5} The dynamics of the charge carriers with momentum
close to the nodal point determine the transport properties in the
normal state. This is particularly the case in the UD region,
where a pseudogap opens along the other directions. In order to
obtain information on the dressing of the charge carriers at the
nodal point, we performed measurements with $\vec{k}$ parallel to
the $(\Gamma-(\pi, \pi))$ direction (see Fig. 2(d)). In Fig. 8 (a)
we show the spectral function  $A(E,k)$ in a false color scale
together with the bare-particle dispersion $\epsilon_k$ \cite
{Kordyuk3}.
\begin{figure}
\centering
\includegraphics[height=7.5cm]{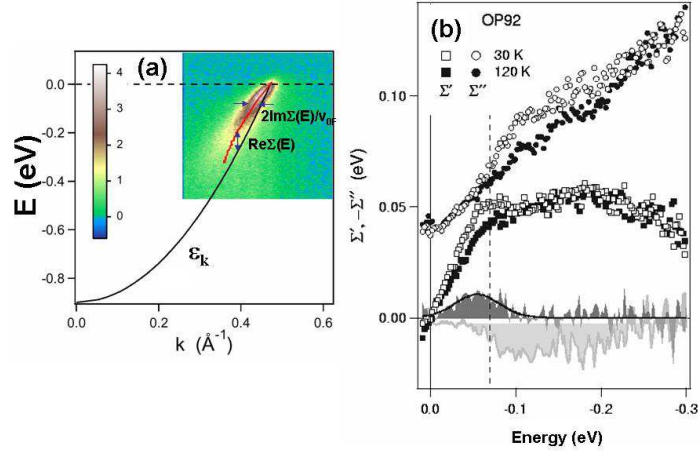}
\caption{ARPES data of optimally doped
(Bi,Pb)$_2$Sr$_2$CaCu$_2$O$_{8+\delta}$ for $\vec{k}$ along to the
nodal direction. (a) spectral function (in false color scale)  at
T = 130 K together with the bare particle dispersion $\varepsilon
_k$ (black line). The red line gives the dispersion derived from
constant E cuts. (b) Real (squares) and negative imaginary
(circles) part of the self-energy function at T = 30 K (open
symbols) and T = 130 K (closed symbols). Dark shaded area:
difference of the real part between the two temperatures. Light
shaded area: negative difference of the imaginary part between the
two temperatures.} \label{fig:8}
\end{figure}
Already without a quantitative analysis, one can learn important
facts from a simple visual inspection of Fig. 8 (a). We clearly
see that there is a strong mass renormalization over an energy
range which extends up to at least 0.4 eV which is much larger
than the energy of the highest phonon modes $E_{ph}$ = 90 meV in
these compounds \cite{Pintschovius}. In these normal state data
the measured dispersion (red line) indicates a "soft" kink at
about 70 meV but comparing the measured spectral function with
that calculated for a single Einstein mode (see Fig. 4 (c)) one
realizes a clear difference. While in the one-mode model there is
a sudden change of the k-dependent width from a resolution
broadened delta-function to a larger width determined by the
constant $\Sigma ''$, in the experimental data there is a
continuous increase of the width (at constant energy) with
increasing binding energy. This clearly excludes the
interpretation in terms of a coupling to a single phonon line and
indicates that the dominant part of the renormalization must be
due to a coupling to an electronic continuum. More information can
be obtained by a quantitative analysis of the data, namely the
extraction of the self-energy function. As described in Sect.
\ref{sec:3.2}, $\Sigma'$ can be derived from the difference
between the bare-particle dispersion and the measured dispersion,
as determined from a fit of the data by a Lorentzian at constant
energy and taking the maximum. From the same fit the width (FWHM)
of the Lorentzian, $\Gamma_k$,  yields $\Sigma ''=\hbar \Gamma_k
v_F/2$.

In Fig. 8 (b) we show $\Sigma'$ and  $\Sigma ''$ of an optimally
doped BiPb2212 crystal measured in the superconducting state at T
= 30 K and in the normal state at T = 130 K \cite{Kordyuk7}. The
data can be analyzed in terms of 3 different scattering channels.
The first channel related to elastic scattering from the potential
of the dopant atoms and possibly also from defects at the surface
can explain about 20 $\%$ the offset of $\Sigma ''$ at zero
energy. The other 80 $\%$ of the offset are due to the finite
momentum resolution. The second scattering channel in the normal
state can be related to a coupling to a continuum of excitations
extending up to about 350 meV. This leads in the normal state to a
marginal Fermi liquid behavior (see Sect. \ref{sec:3.2}): an
almost linear energy dependence of the scattering rate and at low
temperatures an energy dependence of $\Sigma '$ close to $ElnE$.
The continuum to which the charge carriers couple has a cut-off
energy for $\Sigma '$ of about 350 meV. It is remarkable that this
energy is close to the energy of twice the exchange integral, J =
180 meV. Assuming a coupling of the charge carriers to magnetic
excitations \cite {Abanov} in a simple approximation
\cite{Chubukov} the self-energy function can be calculated by a
convolution of the bare particle Greens function G$_0$ and the
energy and momentum dependent magnetic susceptibility $\chi$. This
means $\Sigma=g^2(G_0\bigotimes\chi)$ where g is a coupling
constant. For a two-dimensional magnet it is expected that $\chi$
extends up to an energy of 2J and therefore if the self-energy is
determined by magnetic excitations also $\Sigma'$ should have a
cutoff at that energy. This would support the interpretation of
the continuum in terms of magnetic excitations. In this context
one should mention recent ARPES measurements of an Fe film on W,
where also a strong renormalization well above the phonon energies
has been detected which was interpreted in terms of a coupling to
magnetic excitations \cite{Schafer}. On the other hand the cutoff
energy in $\Sigma '$  may be also related to the finite width of
the Cu-O band.

The third scattering channel exists mainly below T$_c$ and its
intensity is getting rather weak at higher temperatures. It causes
a peak in $\Sigma '$ near 70 meV and an edge in $\Sigma ''$ at
about the same energy. This leads to a pronounced change of the
dispersion at the nodal point at $\sim$ 70 meV which was
previously termed the "kink" \cite{Bogdanov}. The differences
between the self-energy functions $\Delta \Sigma '$  and
$\Delta\Sigma ''$ when going from 30 K to 130 K are plotted in
Fig. 8(b) by shaded areas. Both are typical of a self-energy
function determined by a single bosonic mode. The energy of the
mode may be either $\sim$ 70 meV, when the nodal point is coupled
to gap-less other nodal states or $\sim$ 40 meV when they are
coupled to states close to the antinodal point which in the
superconducting state have a gap of 30 meV. A bosonic mode near 40
meV can be related to the magnetic resonance mode, first detected
by inelastic neutron scattering experiments \cite {neutron}, a
collective mode (spin exciton) which is formed inside the spin gap
of 2$\Delta$ and which decays into single particle excitations
above T$_c$ because of the closing of the gap. The mode energy
$\Omega_0$= 40 meV together with a gap energy $\Delta $= 30 meV
yields a kink energy of 70 meV thus explaining the kink by a
coupling of the antinodal point to the nodal point. Previous
ARPES, optical, and theoretical studies \cite {Johnson, Hwang,
Eschrig} have been interpreted in terms of this magnetic resonance
mode. On the other hand, theoretical work \cite{Abanov2} has
pointed out that because of kinematic constraints a coupling of
the antinodal point to the nodal point via the 40 meV magnetic
resonance mode should not be possible. Recently a new magnetic
resonance mode (the Q* mode) near 60 meV has been detected \cite
{Pailhes,Eremin} which may explain the above mentioned coupling
between nodal points.

In principle the appearance of a sharper kink in the
superconducting state and a {\em decrease} of the scattering rate
in the superconducting state \cite{Kaminski} has been also
explained by the opening of a superconducting gap in the continuum
\cite {Chubukov}. On the other hand, the data shown in Fig. 8 (b)
could indicate that in the superconducting state when compared
with the normal state, there is an {\em additional} scattering
channel and not a {\em reduction} of the scattering rate.

\begin{figure}
\centering
\includegraphics[height=7cm]{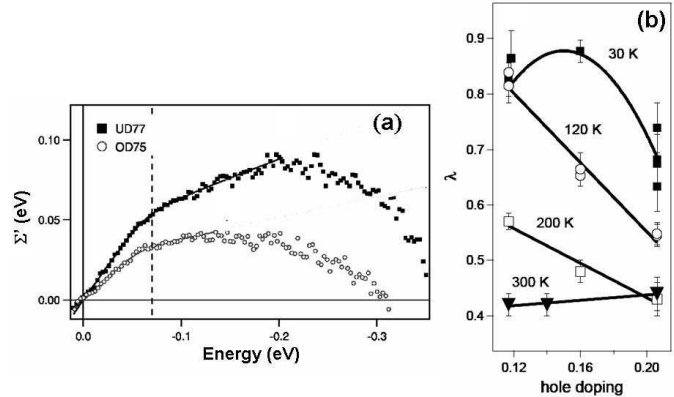}
\caption{ (a)Real part of the self-energy function, $\Sigma '$,
for two (Bi,Pb)$_2$Sr$_2$CaCu$_2$O$_{8+\delta} $ samples at T =
130 K at the nodal point. UD77: underdoped with T$_c$=77 K, OD75:
overdoped with T$_c$=75 K. (b) coupling constant $\lambda$ at the
nodal point as a function of hole concentration for various
temperatures} \label{fig:9}
\end{figure}

In the following we discuss the doping and temperature dependence
of the renormalization effects at the nodal point. In Fig. 9(a) we
show the doping dependence of the real part of the self-energy
function above T$_c$ \cite {Kordyuk3}. Here the contributions from
the third scattering channel, the coupling to a single bosonic
mode, have almost disappeared and mainly a coupling to the
continuum is observed. A rather strong doping dependence is
realized. In the UD sample $\Sigma '$ is much larger and extends
to much higher energies compared to the OD sample. This could
support the assumption that the continuum is related to magnetic
excitations, which increase when approaching the Mott-Hubbard
insulator. From the slope at zero energy (see Sect.
\ref{sec:3.2}), $\lambda$ values could be derived which are
summarized in Fig. 9 (b). The strong doping dependence of
$\lambda$ in the normal state questions the postulation that
independent of the dopant concentration there is a universal Fermi
velocity \cite {Zhou}.

In the normal state $\lambda$ decreases with increasing hole
concentration and increasing temperature. This is expected in the
scenario of a coupling to a continuum of overdamped spin
excitations since for the susceptibility of these excitations a
similar doping and temperature dependence is expected. At 300 K
$\lambda$ is almost independent of the hole concentration.
Possibly there the contribution from the coupling to a continuum
of magnetic excitations has become smaller than the contributions
from electron-hole excitations without spin reversal. The
temperature dependence of the coupling constant at lower hole
concentrations is consistent with the marginal Fermi liquid model,
since there at high temperatures the low-energy properties are no
more determined by the energy dependence and therefore $\lambda$
should decrease with increasing temperature. This is in stark
contrast to the normal Fermi liquid behavior which is observed in
the OD sample (see below).

In the superconducting state there is an additional increase of
$\lambda$, the concentration dependence of which is quite
different from that in the normal state. This clearly indicates
once more the existence of a new additional scattering channel
below T$_c$.

The scattering rate being linear in energy for the OP doped sample
at 130 K transforms {\em continuously} into a more quadratic one
both in the normal and the superconducting state \cite
{Koitzsch2,Kordyuk4}. This indicates that both the second and the
third scattering channel decrease with increasing hole doping,
which is expected in the magnetic scenario. The doping dependence
shows in the normal state a transition from a marginal Fermi
liquid behavior to a more normal Fermi liquid behavior at high
hole concentrations. The quadratic increase in energy (see Sect.
~\ref{sec:3.2})  of $\Sigma '$ is determined by the coefficient
$\beta $ = 1.8 (eV)$^{-1}$. This coefficient is much larger than
the value 0.14 derived for electrons forming the Mo(110) surface
states \cite {Valla2}, This indicates that even in OD HTSCs
correlation effects are still important and electron-electron
interactions and possibly still the coupling to spin fluctuations
are strong.

The strong doping and temperature dependence of the additional
(bosonic) channel is difficult to explain in terms of phonon
excitations. We therefore offered for the additional third
scattering channel an explanation in terms of a coupling to a
magnetic neutron resonance mode, which only occurs below T$_c$.
Finally we mention that an explanation of the extension of the
renormalization to high energies in terms of a multi-bosonic
excitation is very unlikely. A $\lambda$ value below 1, which
corresponds to a quasi-particle spectral weight $Z^{-1}$ larger
than 0.5 would not match with a coupling to polaronic
multi-bosonic excitations.
\section{The dressing of the charge carriers at the antinodal point}
\label{sec:6} Most of the ARPES studies in the past were focused
on the nodal point, where narrow features in  $(E,k)$ space have
been detected, indicating the existence of quasiparticles far down
in the underdoped or even slightly doped region. On the other hand
the antinodal point is of particular interest concerning the
superconducting properties, since in the d-wave superconductors
the superconducting order parameter has a maximum at the antinodal
point \cite{Ding}. The region near the $(\pi ,0)$ point has been
always much more difficult to investigate due to complications of
the bilayer splitting, which could not be resolved by ARPES for 15
years. On the other hand, as mentioned above, only with bilayer
systems of the Bi-HTSC family the entire superconducting range
from the UD to the OD region can be studied. Thus due to the
existence of two Fermi surfaces and two bands close to the Fermi
level near $(\pi ,0)$, with a reduced resolution only a broad
distribution of spectral weight could be observed, leading to the
conclusion that in this $(E,k)$ range very strong interactions
appear causing a complete incoherence of the dynamics of charge
carriers \cite{Shen}. Moreover, in the superconducting state, very
early a peak-dip-hump structure has been observed for all dopant
concentrations which in analogy to the tunnelling spectra in
conventional superconductors, was interpreted as a strong coupling
to a bosonic excitation \cite{Shen}. This picture partially
changed with the advent of the improved experimental situation.

First of all it has been shown by photon-energy dependent
measurements in the range $h\nu$ = 20 - 60 eV using synchrotron
radiation \cite {Borisenko3, Kordyuk5} that the peak-dip-hump
structures strongly change as a function of the photon energy.
This indicated that the matrix element  in Eq. (5) has a different
photon energy dependence for  the bonding and the antibonding band
at $(\pi ,0)$. This experimental observation was confirmed by
calculations of the matrix element using LDA bandstructure
calculations \cite{Bansil, Fujimori}. It turned out that the
peak-dip-hump structure in the OD sample was dominated by the
bilayer splitting, i.e, the peak is caused by the antibonding band
and a hump is caused by the bonding band. In the UD range the
complicated spectral shape could be traced back to a superposition
of the bilayer effects and strong renormalization effects in the
superconducting state. In this situation, only momentum dependent
measurements \cite{Kim,Gromko,Sato} along the
$(\pi,\pi)-(\pi,-\pi)$ line could separate the two effects. In
Fig. 10 a collection of our ARPES data along this direction,
centered around the $(\pi,0)$ point, is shown as a function of the
dopant concentration in the superconducting state (T = 30 K). In
the lowest row, normal state data (T = 120K) are also shown for
the UD sample.
\begin{figure}
\centering
\includegraphics[height=10cm]{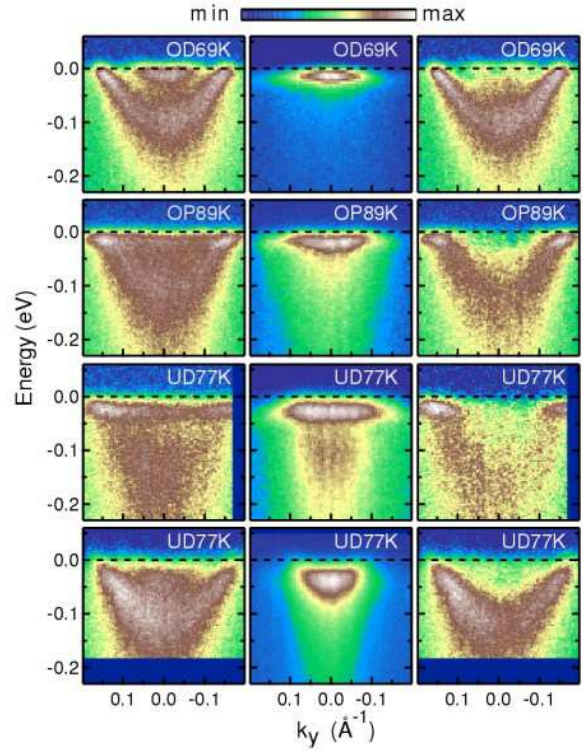}
\caption{ARPES intensity plots as a function of energy and wave
vector along the $(\pi,\pi)-(\pi,-\pi)$ direction of overdoped
(OD), optimally doped (OP) and underdoped  (UD)
(Bi,Pb)$_2$Sr$_2$CaCu$_2$O$_{8+\delta} $ superconductors taken at
T = 30 K (upper 3 rows). Zero  corresponds to the $(\pi,0)$ point.
Fourth row: data for an UD sample taken at T = 120 K. Left column:
data taken with a photon energy  $h\nu$=38 eV, at which the signal
from the bonding band is maximal. Middle column: data taken at
h$\nu$=50 eV (or 55 eV), where the signal from the antibonding
band is dominant. Right column: subtraction of the latter from the
former yielding the spectral weight of the bonding band.}
\label{fig:10}
\end{figure}
In the upper left corner the data for an OD sample clearly show
the splitting into a bonding and an antibonding band related to
four Fermi surface crossings and two saddle points as expected
from the tight-binding bandstructure calculations, shown in Fig. 2
(a) taking into account the bilayer splitting  visualized in Fig.
2 (d). Looking in the same column at the  low temperature data of
the OP and UD sample the two bands are no more resolved. As
mentioned before the matrix element for the excitation of the 2
bands is strongly photon energy dependent and it was shown \cite
{Bansil, Fujimori, Borisenko3, Kordyuk5} that the spectra in the
first column which were taken at $h\nu$ = 38 eV represent mainly
the bonding  band with some contributions from the antibonding
band. The data in the second column were taken with $h\nu$ = 50
(or 55) eV  and have almost pure antibonding character.
Subtracting the second column from the first column yields almost
the pure spectral weight from the bonding band. Using this
procedure one clearly recognizes that  even in the UD samples the
bonding and the antibonding band can be well separated. In the
superconducting state (first 3 rows) these data show  strong
changes upon reducing the dopant concentration. The bonding, and
most clearly seen, the antibonding band move further and further
below the Fermi level, indicating the reduction of holes. In the
bonding band of the OD crystal almost no kink is observed but in
the OP  sample a very strong kink  is realized, disclosed by the
appearance of a flat dispersion between the gap energy at about
$\sim $ 30 meV and the branching energy of $\sim $70 meV followed
by a steeper dispersion and a strong broadening. The strong
renormalization effects increase even further when going from the
OP doped sample to the UD sample. Remarkably, the renormalization
effects (with the exception of the pseudogap) described above,
completely disappear in the normal state as can be seen in the
fourth row where data from an UD sample, taken at 120 K, are
shown.  As in the OD sample, a normal dispersion without a kink is
now detected for the bonding band. Also for the antibonding band
there is a transition from a flat band at low temperatures to a
dispersive band above T$_c$. A comparison with the bare-particle
band structure (not shown) indicates that there is reduction of
the bandwidth by a factor of about 2 which means that there is a
$\lambda^w$ of about 1 in the normal state. A renormalization
corresponding to a $\lambda^w$ of about 1 is also detected above
the branching energy near $(\pi ,0)$ in the superconducting state.
This bandwidth renormalization in the range of the antinodal point
is similar to that one at the nodal point and is probably also
related to a coupling to  a continuum of magnetic excitations.

In Fig. 11 (a) we show an ARPES intensity distribution of the
antibonding band near $k_F$ measured with a photon energy h$\nu$ =
50 eV along the $(1.4\pi,\pi)-(1.4\pi,-\pi)$ line for OP Pb-Bi2212
at 30 K. At this place in the second Brillouin zone the bare
particle dispersion of the antibonding band reaches well below the
branching energy $E_B$ = 70 meV. Therefore, contrary to the data
shown in Fig. 10 (second row, second column), which were taken
along the $(\pi,\pi)-(\pi,-\pi)$ line, the branching into two
dispersive arms can be clearly realized. The data in Fig. 11 (a)
together with those shown in Fig. 10 (second row, third column)
for the bonding band, when compared with the model calculations
shown in Fig. 5, clearly show that the dominant renormalization
effect in the superconducting state is a coupling to a bosonic
mode \cite{ Kaminski, Norman}.

To obtain more information about the renormalization and the
character of the mode,  the spectral function was analyzed
quantitatively \cite{Fink}.
\begin{figure}
\centering
\includegraphics[height=6.5cm]{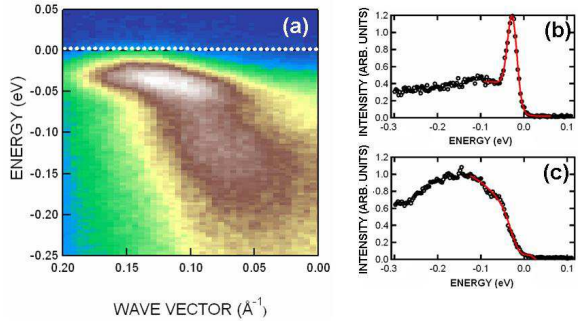}
\caption{(a) Spectral function for k-values near the
$(1.4\pi,\pi)-(1.4\pi,-\pi)$ direction  of the OP
(Bi,Pb)$_2$Sr$_2$CaCu$_2$O$_{8+\delta} $ superconductor taken at T
= 30 K . Zero  corresponds to the $(1.4\pi,0)$ point. The data
were taken with a photon energy  h$\nu$ = 50 eV in order to
maximize the intensity of the antibonding band. (b) Constant-k cut
of the spectral weight of the bonding band (see Fig. 10, optimally
doped sample, T = 30 K) at k$_F$. (c) Cut of the data at about one
third of k$_F$ (starting from $(\pi ,0)$).}
 \label{fig:11}
\end{figure}
Cutting the measured intensity distribution of the bonding band
(see Fig. 10) at k$_F$ yields the peak-dip-hump structure shown in
Fig. 11 (b). From the peak energy one can derive a superconducting
gap energy of $\Delta$ = 30 meV. Cutting the data at about 1/3 of
k$_F$ (starting from $(\pi ,0)$) yields the spectrum shown in Fig.
11 (c). There the coherent peak is strongly reduced and in the
framework of  a one-mode model, the threshold after the coherent
peak, derived from a fit of the spectrum, yields the branching
energy $\Omega _0 + \Delta $ = 70 meV. Another way to obtain the
branching energy is to determine the threshold of $\Sigma ''$
which can be obtained by fitting the spectral weight of constant
energy cuts using Eq. (10). From this, the branching energy
$\Omega _0+\Delta$ = 70 meV can be derived. From fits of constant
energy cuts just below the branching energy the parameter $\Sigma
''(-\infty) \sim$ 130 meV can be obtained which is also a
measurement of the coupling of the charge carriers to a bosonic
mode (see Sect.~\ref{sec:3.2}).

Important information comes from the dispersion between the gap
energy and the branching energy. Originally \cite {Kim} the data
were fitted using Eq. (12) yielding $\lambda$ values as a function
\begin{figure}
\centering
\includegraphics[height=5cm]{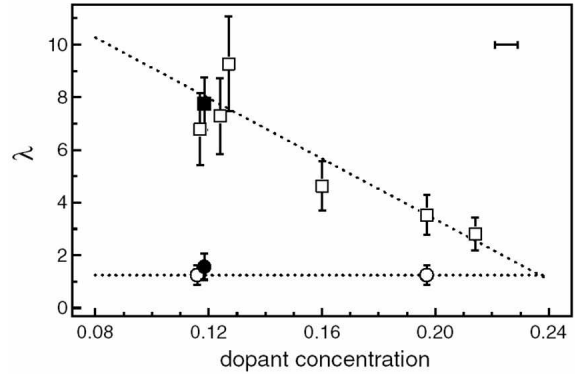}
\caption{The coupling strength parameter $\lambda $ at the
antinodal point as a function of doping concentration. Squares:
superconducting state; circles: normal state; open (solid)
symbols: bonding (antibonding) band.}.
 \label{fig:12}
\end{figure}
of the dopant concentration shown in Fig.~12. However, as pointed
out in Sect.~\ref{sec:3.3},  the situation in HTSCs is quite
different from conventional superconductors. In the former $\Omega
_0$ is not much larger than $\Delta$  and therefore the function
$Z(E)$, from which  $\lambda $ is derived, is energy dependent.
Furthermore, as shown in Sect.~\ref{sec:3.3} the $\lambda$ values,
evaluated in this way, depend on the gap energy. In the one-mode
model the gap energy dependence of $\lambda$ is determined  by the
factor $(\Omega _0+\Delta )/\Omega _0$. Therefore it is
questionable whether those $\lambda$ values are  a good measure of
the coupling strength to a bosonic mode. More recently
\cite{Fink}, we have fitted the dispersion of the coherent peak of
an OP doped sample using the full Eq. (11) taking into account the
above mentioned band renormalization by a factor of 2 using a
$\lambda^w \sim 1$. From the derived $Z(E)$ in the superconducting
state, $Z(E)$ in the normal state could be calculated by setting
$\Delta$ to zero and then a total coupling constant $\lambda ^t_n$
= 3.9 could be obtained which is composed of a $\lambda ^b_n$ =
2.6 due to the coupling to the bosonic mode and a $\lambda^w_n$ =
1.3 from the band renormalization. It is interesting that this
value is close to the value derived using Eq. (12). The reason for
this is that the reduction due to the energy dependence of Z is
partially compensated by the transformation into the normal state.
A first estimate shows that the values at other dopant
concentrations are also not drastically changed.

One may argue that those very large $\lambda $-values are
unphysical and meaningless because, in the case of electron-phonon
coupling, lattice instabilities may be expected. On the other hand
also another measure of the coupling strength, the imaginary part
of the self-energy function above the branching energy is very
large. From  $\Sigma ''(-\infty)$ = 130 meV and $\Omega _0$ = 40
meV one obtains (see Sect. \ref{sec:3.2}) $\lambda = \Sigma
''(-\infty)/(\pi \Omega _0/2)$ = 2.1 which is not far from the
above given value $\lambda ^b_n$ = 2.6 for the coupling to the
bosonic mode derived from the dispersion.

It is interesting to compare the present values  of $\lambda ^b_n$
= 2.6 and $\Sigma ''(-\infty)$ = 130 meV derived for an OP HTSC in
the superconducting state with those obtained for the
electron-phonon coupling of surface electrons on a Mo(110) surface
($\lambda ^b_n$ = 0.42 and $\Sigma ''(-\infty)_{el-ph}$ = 30 meV)
\cite{Valla}. So both parameters are for the HTSC a factor 4-6
larger than for the Mo(110) surface. This indicates that in the
HTSCs in the superconducting state there is really an anomalous
strong coupling to a bosonic mode, which manifests itself both in
the high coupling constant and in the high scattering rates above
the branching energy. Finally it is remarkable that there is no
indication of a multibosonic excitation comparable to that in the
undoped cuprates \cite {KShen,Roesch} since in that case, taking
the above derived $\lambda^b$ values, the intensity of the
coherent state relative to the incoherent states should be
strongly reduced in disagreement with the data shown in Figs. 10
and 11.
\begin{figure}
\centering
\includegraphics[height=4cm]{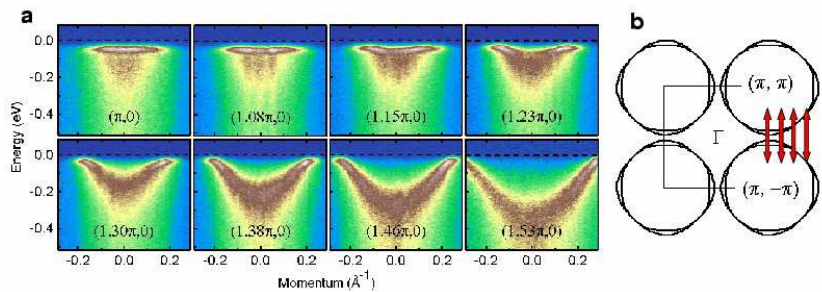}
\caption{Intensity distribution for cuts in the Brillouin zone indicated in the right-hand sketch of the optimally doped
(Bi,Pb)$_2$Sr$_2$CaCu$_2$O$_{8+\delta} $ superconductor taken at T = 30 K .
 Upper left panel: antinodal point. Lower right panel: nodal point.
The data were taken with a photon energy  h$\nu$=50 eV in order to maximize the intensity of the
antibonding band.}.
 \label{fig:13}
\end{figure}

In Fig. 13 we show the renormalization of the antibonding band in
the superconducting state for an OP sample when going from the
antinodal point to the nodal point \cite{thesis}. By looking at
the dispersion close to E$_F$ the renormalization is large at the
antinodal point and it is much weaker at the nodal point. This is
in line with a strong coupling to a mode which is related to a
high susceptibility for a wave vector $(\pi ,\pi )$ which leads to
a coupling between antinodal points.

This leads to the above mentioned spin fluctuation scenario, in
which below T$_c$, the opening of the gap causes via a feed-back
process the appearance of a magnetic resonance mode, detected by
inelastic neutron scattering \cite{neutron}. This mode has a high
spin susceptibility at the wave vector $(\pi ,\pi )$,  the energy
is $\sim$ 40 meV, and  as mentioned above it exists only below
T$_c$. Thus from the measurements of the spectral function around
($\pi $,0), in particular from the energy, the momentum, and the
temperature dependence we conclude that the mode to which the
charge carriers at  the antinodal point so strongly couple, is the
magnetic resonance mode. In a  recent theoretical work
\cite{Eschrig} it was pointed out that according to magnetic
susceptibility measurements using inelastic neutron scattering the
magnetic resonance mode couples the antibonding band predominantly
to the bonding  band and vice versa. This means only the odd
susceptibilities ${\chi_{AB}}$ and ${\chi_{BA}}$ are large and the
even susceptibilities ${\chi_{AA}}$ and ${\chi_{BB}}$ are small.
There is no coupling via the resonance mode within a band. It is
remarkable that the coupling of the bonding band to the resonance
mode starts in the OD region near 22 \% doping when the saddle
point of the antibonding band just crosses the Fermi level (see
Fig. 12). The result that in the UD region $\lambda$ is similar
for both bands is understandable, since the Fermi velocities and
therefore the density of states and the odd susceptibilities
${\chi_{AB}}$ and ${\chi_{BA}}$ should be comparable. This
scenario is supported by recent measurements of the energy
dependence of the different scattering rates of the bonding and
the antibonding band close to the nodal point \cite{Borisenko5}.
Similar data have been recently presented for the system
YBa$_2$Cu$_3$O$_7$ \cite{Borisenko6}. Furthermore, recently our
group has observed large changes of the renormalization effects at
the nodal and the antinodal point upon substituting 1 or 2 $\%$ of
the Cu ions by nonmagnetic Zn (S=0) or magnetic Ni (S=1),
respectively \cite{Zabolotnyy}. These strong changes also strongly
support the magnetic scenario since this substitution of a very
small amount of the Cu ions should not change the coupling of the
charge carriers to phonons. On the other hand we do not want to
conceal that there are also interpretations of the above discussed
bosonic mode in terms of phonon excitations \cite {Devereaux}.

At the end of this Section we would like to mention some ARPES
results on the spectral function in the pseudogap region \cite
{Eckl}. The pseudogap is one of the most remarkable properties of
HTSCs in the UD region above T$_c$. In Fig. 14 we compare the
dispersion along the $(\pi ,0)-(\pi ,-\pi )$ direction close to
the antinodal point of an UD sample in the superconducting  and in
the pseudogap state.
\begin{figure}
\centering
\includegraphics[height=4cm]{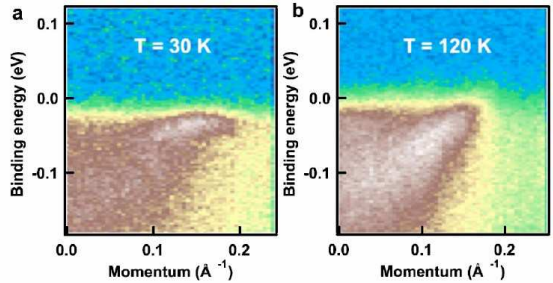}
\caption{Intensity distribution near the  antinodal point along the $(\pi,0)-(\pi,-\pi)$ direction of an underdoped
(Bi,Pb)$_2$Sr$_2$CaCu$_2$O$_{8+\delta} $ crystal with T$_c$ =77 K. The photon energy has been
chosen to be 38 eV in order to suppress the antibonding band. (a) superconducting state.
(b) pseudogap state.}
 \label{fig:14}
\end{figure}
In the superconducting state one realizes the characteristic
BCS-Bogoliubov-like back-dispersion at k$_F$. In the pseudogap
state no more a bending back of the dispersion is observed.
Instead the spectral weight fades when the binding energy
approaches the gap energy. This is in line with the observed
disappearance of the coherent peak in tunnelling spectra of HTSCs
in the pseudogap phase \cite {Kugler}. The experimental
observation of this behavior was explained in terms of phase
fluctuations of the superconducting order parameter. There is a
general uncertainty relation for superconductors for the particle
number $N$ and the phase $\phi $, $\Delta N\Delta \phi \geq 1$. In
the superconducting state at T = 0 with $\Delta \phi $ = 0, the
particle number is completely uncertain leading to a large
particle-hole mixing and thus to a large back-dispersion. With
increasing temperature, the phases get completely uncorrelated and
one obtains $\Delta N$ = 0. Then the back-dispersion must
disappear. In this way a crossover from a BCS-like phase-ordered
bandstructure to a completely new phase-disordered pseudogap
bandstructure is obtained.
\section{Conclusions}
\label{sec:7} In this chapter we presented part of our ARPES
results on HTSCs. They were obtained with an energy and a momentum
resolution of 8-25 meV and 0.015 \AA$^{-1}$. This is at present
the state of the art ARPES when reasonable intensities are used
during the measurements. Using this resolution, a lot of
information on the dressing of the charge carriers in HTSCs at
different k-points in the Brillouin zone has been obtained during
the last 10 years.  It will be really one of the big challenges of
experimental solid state physics to enter in the range of
sub-1meV-resolution in {\em angle-resolved} photoemission
spectroscopy. In the next 10 years it is predictable that there
will be a further improvement of the energy resolution by one
order of magnitude for angle-resolved measurements. It can be
anticipated that further interesting results on the dressing and
possibly also on the pairing mechanism in HTSCs can be realized.
\vspace{2.0em}

\index{Acknowledgments}
{\bf Acknowledgements}

We acknowledge financial support by the DFG Forschergruppe under Grant No. FOR 538. One of the authors (J.F.)
appreciates the hospitality during his stay at the Ames Lab and thanks for the critical reading of the manuscript
by D. Lynch. We thank R. Follath, T.K. Kim, S. Legner, and K.A. Nenkov for  fruitful collaboration.
In particular we thank Mark Golden for his contributions in the early stage of the project. Finally we acknowledge helpful
discussions and collaboration with colleagues from theory: A.V. Chubukov, T. Eckl, M. Eschrig, and W. Hanke.

%
\bibliographystyle{unsrt}
\bibliography{}

\begin{thebibliography}{99.}
\bibitem{Einstein} A. Einstein: Ann. d. Phys.
\textbf{31}, 132 (1905)
\bibitem{Hertz} H. Hertz: Ann. d. Phys.
\textbf{17}, 983 (1887)
\bibitem{Gobeli} G.W. Gobeli, F.G. Allen, E.O. Kane: Phys. Rev. Lett.
\textbf{12}, 94 (1964)
\bibitem{Muller} G. Bednorz, K.A. M\"uller: Z. Phys. B
\textbf{64}, 189 (1986)
\bibitem{Zaanen} J. Zaanen, G.A. Sawatzky, J. Allen: Phys. Rev. Lett.
\textbf{64}, 189 (1986)
\bibitem{Lynch} D.W. Lynch, C.G. Olson: \textit{Photoemission Studies of High-Temperature Superconductors},
(Cambridge University Press 1999) pp 1-432
\bibitem{Damascelli} A. Damascelli, Z.-X. Shen, Z. Hussain: Rev. Mod. Phys.
\textbf{75}, 473 (2003)
\bibitem{Campuzano} J.C. Campuzano, M.R. Norman, M. Randeria: Photoemission in the High-T$_c$
Superconductors. In: \textit{Physics of Superconductors}, vol II, ed by
K.H. Bennemann and J.B. Ketterson (Springer, Berlin Heidelberg New York 2004) pp
167-273
\bibitem{Timusk} T. Timusk: Rep. Progr. Phys. {\bf 62}, 61 (1999)
\textbf{75}, 473 (2003)
\bibitem{Varma2}C.M. Varma: Phys. Rev. B {\bf 55}, 14554 (1997)
\bibitem{Andersen}O.K. Andersen, A.I. Liechtenstein, O. Jepsen, F. Paulsen:
J. Phys. Chem. Solids {\bf 56}, 1573 (1995)
\bibitem{Ding} H. Ding, M.R. Norman, J.C. Campuzano, M. Randeria, A. Bellman,
T. Yokoya, T. Takahashi, T. Mochiku, K. Kadowaki: Phys. Rev. B {\bf 54}, R9678 (1996)
\bibitem{Borisenko2}S.V. Borisenko, A.A. Kordyuk, T.K. Kim, S. Legner, A. Nenko,  M. Knupfer,  M.S. Golden, J. Fink,
 H. Berger, R. Follath: Phys. Rev. B. {\bf 66}, 140509(R) (2002)
\bibitem{Kordyuk6}A.A. Kordyuk, S.V. Borisenko, A.N. Yaresko, S.-L. Drechler, H. Rosner,
T.K. Kim, A. Koitzsch, K.A. Nenkov, M. Knupfer, J. Fink, R.
Follath, H. Berger, B. Keimer, S. Ono, Y. Ando: Phys. Rev. B {\bf
70}, 214525 (2004)
\bibitem{Nucker1}N. N\"ucker, J. Fink, J.C. Fuggle, P.J. Durham, W. Temmerman: Phys. Rev. B {\bf 37}, 6827 (1988)
\bibitem{Hufner} S. H\"ufner: \textit{Photoelectron Spektroscopy},
 (Springer, Berlin Heidelberg New York 1996)and references therein.
\bibitem{Mahan} G.D. Mahan: \textit{Many-Particle Physics},
 (Plenum Press, New York 1990)
\bibitem{Hedin} L. Hedin, S. Lundquist: Solid State Physics
\textbf{23}, 1963 (1969)
\bibitem{Almbladh} C.O. Almbladh, L. Hedin: Beyond the one-electron model
 in \textit{Handbook of Synchrotron Radiation},
vol 1b, ed by E.E. Koch (North Holland, Amsterdam 1983) pp. 607-904.
 \bibitem{Varma}C.M. Varma, P.B. Littlewood, S. Schmitt-Rink, E. Abrahams, A.E. Ruckenstein:
Phys. Rev. Lett. {\bf 63}, 1996 (1989)
\bibitem{Pines} D. Pines, P. Nozieres: \textit{The Theory of Quantum Liquids},vol 1,
(W.A. Benjamin, New York 1966) p. 64
\bibitem{Imada} M. Imada, A. Fujimori, Y. Tokura: Rev. Mod. Phys. {\bf 70}, 1039 (1998)
\bibitem{Valla}T. Valla, A.V. Federov, P.D. Johnson, B.O. Wells,
S.L. Hulbert, Q. Li, G.D. Gu, N. Koshizuka: Science {\bf 285}, 2110 (1999)
\bibitem{Hodges}C. Hodges, H. Smith, J.W. Wilkins: Phys. Rev. B {\bf 4}, 302 (1971)
\bibitem{Schrieffer} S. Engelsberg, J. R. Schrieffer: Phys. Rev. {\bf 131}, 993 (1963)
\bibitem{KShen} K.M. Shen, F. Ronning, D.H. Lu, W.S. Lee, N.J.C. Ingle, W. Meevasana, F. Baumberger, A. Damascelli,
N.P. Armitage, L.L. Miller, Y. Kohsaka, M. Azuma, M. Takano, H. Takagi, Z.-X. Shen :
Phys. Rev. Lett. {\bf 93}, 267002 (2004).
\bibitem{Roesch} O. R\"osch, O. Gunnarsson:
Phys. Rev. Lett. {\bf 92}, 146403 (2004).
\bibitem{Nambu} Y. Nambu:
Phys. Rev. {\bf 117}, 648 (1960).
\bibitem{Scalapino} D. J. Scalapino, in  \textit{Superconductivity}, vol. 1,
ed. R. D. Parks (Marcel Decker, New York 1969),  p. 449.
\bibitem{Chubukov} A.V. Chubukov, M.R. Norman:
Phys. Rev. B {\bf 70}, 174505 (2004).
\bibitem{Martensson} N. Martensson, P. Baltzer, P. A. Bruhwiler,
J.-O. Forsell, A. Nilsson, A. Stenborg, B. Wannberg: Journal of Electron
Spectroscopy and Related Phenomena {\bf 70}, 117 (1994)
\bibitem{Koch} \textit{Handbook on Synchrotron Radiation}, vol 1-4, ed by
E.E. Koch, E.V. Marr, G.S. Brown, D.E. Moncton, S. Ebashi, M. Koch, E. Rubenstein
 (North Holland, Amsterdam 1983-1991)
\bibitem{Follath}R. Follath: Nucl. Instrum. Meth, Phys. Res. A {\bf 467-468}, 418 (2001)
\bibitem{Borisenko1}S.V. Borisenko, M.S. Golden, S. Legner, T. Pichler, C. D\"urr,  M. Knupfer, J. Fink,
G. Yang, S. Abell,  H. Berger: Phys. Rev. Lett. {\bf 84}, 4453 (2000)
\bibitem{Legner}S. Legner, S.V. Borisenko, C. D\"urr, T. Pichler, M. Knupfer,
M.S. Golden, J. Fink, G. Yang, S. Abell, H. Berger,
R. M\"uller, C. Janowitz, G. Reichardt: Phys. Rev. B {\bf 62}, 154 (2000)
\bibitem{Kordyuk1}A. A. Kordyuk, S. V. Borisenko, M.S. Golden, S. Legner, K.A. Nenkov,
M. Knupfer, J. Fink, H. Berger, L. Forro, R. Follath: Phys. Rev. B {\bf 66}, 014502 (2002)
\bibitem{Tallon}J.L. Tallon, C. Bernhard, H. Shaked, R.L. Hitterman, J. D. Jorgensen:
Phys. Rev. B {\bf 51}, 12911 (1995)
\bibitem{Feng} D.L. Feng, N.P. Armitage, D.H. Lu, A. Damascelli, J.P. Hu, P. Bogdanov,
A. Lanzara, F. Ronning, K.M. Shen, J.-I. Shimoyama, K. Kishio: Phys. Rev. Lett.
{\bf 86}, 5550 (2001)
\bibitem{Chuang}Y.-D. Chuang, A.D. Gromko, A. Federov, Y. Aiura, K. Oka,
Yoichi Ando, H. Eisaki, S. L. Uchida, D.S. Dessau: Phys. Rev. Lett.{\bf 87}, 117002 (2001)
\bibitem{Bansil} A. Bansil, M. Lindroos, Phys. Rev. Lett. {\bf 83}, 5154 (1999)
\bibitem{Fujimori}J.D. Lee and A. Fujimori, Phys. Rev. Lett. {\bf87}, 167008 (2001)
\bibitem{Aebi}P. Aebi, J. Osterwalder, P. Schwaller, L. Schlapbach,
M. Shimeda, T. Muchiku, K. Kadawaki: Phys. Rev. Lett. {\bf 72}, 2757 (1994)
\bibitem{Koitzsch1}A. Koitzsch, S.V. Borisenko, A.A. Kordyuk, T.K. Kim, M. Knupfer, J. Fink,
M.S. Golden, W. Koops, H. Berger, B. Keimer, C.T. Lin, S,. Ono, Y.
Ando, R. Follath: Phys. Rev. B {\bf 69}, 220505(R) (2004)
\bibitem{Kordyuk2}AA. Kordyuk, S.V. Borisenko, M. Knupfer, J. Fink: Phys. Rev. B
{\bf 67}, 064504 (2003)
\bibitem{Nucker}N. N\"ucker, U. Eckern, J. Fink, P. M\"uller: Phys. Rev. B {\bf 44}, 7155 (1991)
\bibitem{Grigoryan}V.G. Grigorian, G. Paasch, S.-L. Drechsler: Phys. Rev. B {\bf 60}, 1340 (1999)
\bibitem{Kordyuk3}A.A. Kordyuk, S.V. Borisenko, A. Koitzsch, J. Fink, M. Knupfer,
H. Berger: Phys. Rev. B {\bf 71}, 214513 (2005)
\bibitem{Pintschovius} L. Pintschovius, W. Reichardt: Neutron Scattering in Layered Copper-Oxide
Superconductors. In: \textit{Physics and Chenistry of Materials with Low Dimensional Structures}, vol 20, ed. by
A. Furrer (Kluwer Academic, Dordrecht, 1998) p.165
\bibitem{Kordyuk7}A.A. Kordyuk, S.V. Borisenko, V.B. Zabolotnyy, J. Gerk, M. Knupfer, J.
Fink, B. B\"uchner, C.T. Lin, B. Keimer, H. Berger, Seiki Komiya,
Yoichi Ando: cond-mat/0510760.
\bibitem{Abanov} A. Abanov, A.V. Chubukov, J Schmalian:
Adv. Phys. {\bf 52}, 119 (2003).
\bibitem{Schafer} J. Sch\"afer, D. Schrupp, Eli Rotenberg, K. Rossnagel, H. Koch, P. Blaha, R. Claessen:
Phys. Rev. Lett. {\bf 02}, 097205 (2005).
\bibitem{Bogdanov}P.V. Bogdanov. A. Lanzara, S.A. Kellar, X.J. Zhou, E.D. Lu, W.J. Zheng, G. Gu, J.-J. Shimoyama,
K. Kishio, H. Ikeda, R. Yoshizaki, Z. Hussain, Z.-X. Shen: Phys.
Rev. Lett. {\bf 85}, 2581 (2000)
\bibitem{neutron} J. Rossat-Mignod, L.P. Regnault, C. Vettier, P. Bourges, P. Burlet, J. Bossy, J.Y. Henry, G. Lapertot:
 Physica C {\bf185-189}, 86 (1991); H. A. Mook, M. Yethiraj,
G. Aeppli, T.E. Mason, T. Armstrong: Phys. Rev. Lett. {\bf70}, 3490 (1993);
H. F. Fong, B. Keimer, P.W. Anderson, D. Reznik, F. Dogan, I.A. Aksay: {\it ibid.} {\bf75}, 316 (1995).
\bibitem{Johnson}P.D. Johnson, T. Valla, A.V. Feferov,  Z. Yusof, B.O. Wells, Q. Li,
A.R. Moodenbaugh, G.D. Gu, N. Koshizuka, C. Kendziora, Sha Jian, D.G. Hinks: Phys. Rev. Lett. {\bf 87}, 177007 (2001)
\bibitem{Hwang}J. Hwang, J. Yang, T. Timusk, S.G. Sharapov, J.P. Carbotte, D.A. Bonn, R. Liang, W.N. Hardy: cond-mat/0505302
\bibitem{Eschrig} M. Eschrig and M. R. Norman: Phys. Rev. Lett. {\bf85}, 3261
(2000), Phys. Rev. Lett. {\bf89}, 277005 (2002), Phys. Rev. B
{\bf67} 144503 (2003).
\bibitem{Abanov2} A. Abanov, A.V. Chubukov, M. Eschrig, M.R. Norman, and J. Schmalian:
Phys. Rev. Lett. {\bf89}, 177002 (2002)
\bibitem{Pailhes} S. Pailhes, Y. Sidis, P. Bourges, V. Hinkov, A.
Ivanov, C. Ulrich, L.P. Reqnault, B. Keimer: Phys. Rev. Lett.{\bf
93}, 167001 (2004)
\bibitem{Eremin}I. Eremin, D. K. Morr, A.V. Chubukov, K. Bennemann, M.R. Norman: cond-mat/0409599.
\bibitem{Kaminski}A. Kaminski, M. Randeria, J.C. Campuzano, M.R. Norman, H. Fretwell,
J. Mesot, T. Sato, T. Takahashi, K. Kadowaki: Phys. Rev. Lett.
{\bf 86}, 1070 (2001)
\bibitem{Zhou}X.J. Zhou, T. Yoshida, A. Lanzara, P.V. Bogdanov, S.A. Kellar,
K.M. Shen, W.L. Yang, F. Ronning, T. Sasagawa, T. Kakeshita, T.
Noda, H. Eisaki, S. Uchida, C.T. Lin, F. Zhou, J.W. Xiong, W.X.
Ti, Z.X. Zhao, A. Fujimori, Z. Hussain, Z.-X. Shen: Nature {\bf
423}, 398 (2003)
\bibitem{Koitzsch2}A. Koitzsch, S.V. Borisenko, A.A. Kordyuk, T.K. Kim, M. Knupfer,
J. Fink, H. Berger, R. Follath: Phys. Rev. B. {\bf 69}, 140507(R)
(2004)
\bibitem{Kordyuk4}A.A. Kordyuk, S.V. Borisenko, A. Koitzsch, J. Fink, M. Knupfer, B. B\"uchner,
H. Berger, G. Margaritondo, C.T. Lin, B. Keimer, S. Otto, Y. Ando: Phys. Rev. Lett {\bf 92}, 257006 (2004)
\bibitem{Valla2}T. Valla, A.V. Feferov, P.D. Johnson, S.L. Hulbert: Phys. Rev. Lett. {\bf 83}, 2085 (1999)
\bibitem{Shen} Z.-X. Shen, J.R. Schrieffer:
Phys. Rev. Lett. {\bf 78}, 1771 (1997).
\bibitem{Borisenko3}S.V. Borisenko, A.A. Kordyuk, T.K. Kim, A. Koitzsch,  M. Knupfer, J. Fink, M.S. Golden,
M. Eschrig, H. Berger, R. Follath: Phys. Rev. Lett. {\bf 90},
207001 (2003)
\bibitem{Kordyuk5}A.A. Kordyuk, S.V. Borisenko, T.K. Kim, K.A. Nenkov, M. Knupfer,
J. Fink, M.S. Golden, H. Berger,R. Follath: Phys. Rev. Lett. {\bf
89}, 077003 (2002)
\bibitem{Kim}T.K. Kim, A.A. Kordyuk, S.V. Borisenko, A. Koitzsch, M. Knupfer, H. Berger, J. Fink:
Phys. Rev. Lett. {\bf 91}, 167002 (2003)
\bibitem{Gromko}A.D. Gromko, A. V. Federov, Y.D. Chuang, J.D. Koralek, Y Aiura, Y. Yamaguchi, K. Oka, Yoichi Ando,
D.S. Dessau: Phys. Rev. B {\bf 68}, 174520 (2003)
\bibitem{Sato}T. Sato, H. Matsui, T. Takahashi, D. Ding, H.-B. Yang, S.-C. Wang, T. Fuji, T. Watanabe, A. Matsuda,
T. Terashima, K. Kadowaki: Phys. Rev. Lett. {\bf 91}, 157003
(2003)
\bibitem{Norman} M.R. Norman, H. Ding:
Phys. Rev. B {\bf 57}, 11111 (1998).
\bibitem{Fink}J. Fink, A. Koitzsch,
J. Geck, V.Zabolotnny, M. Knupfer, B. B\"uchner, H. Berger:
cond-mat/0604665
\bibitem{thesis} T.K. Kim: The role of inter-plane interactions in the electronic structure of high-T$_c$
cuprates. PhD Thesis, University of Technology, Dresden (2003)
\bibitem{Borisenko5}S.V. Borisenko, A.A. Kordyuk, A. Koitzsch, J. Fink, J. Geck, V. Zabolotnyy,
M. Knupfer, B\"uchner, H. Berger, M. Falub, M. Shi, J. Krempasky,
L. Patthey: Phys. Rev. Lett. {\bf 96}, 067001 (2006).
\bibitem{Borisenko6}S.V. Borisenko, A.A. Kordyuk, V. Zabolotnyy,
J. Geck, D. Inosov, A. Koitzsch, J. Fink, M. Knupfer, B\"uchner,
V. Hinkov, C.T. Lin, B. Keimer, T. Wolf, S.G. Chiuzbaian, L.
Patthey, R. Follath: Phys. Rev. Lett. {\bf 96}, 117004 (2006).
\bibitem{Zabolotnyy}V. Zabolotnyy, S.V. Borisenko, A.A. Kordyuk, J. Fink, J. Geck, A. Koitzsch, M. Knupfer, B. Büchner,
H. Berger, A. Erb, C.T. Lin, B. Keimer, R. Follath: Phys. Rev.
Lett. {\bf 96}, 037003 (2006).
\bibitem{Devereaux} T.P. Devereaux, T. Cuk, Z.-X. Shen, N. Nagaosa:
Phys. Rev. Lett. {\bf 93}, 117004 (2004).
\bibitem{Eckl} T. Eckl, W. Hanke, S.V. Borisenko, A.A. Kordyuk, T.Kim, A. Koitzsch, M. Knupfer, J. Fink:
Phys. Rev. B {\bf 70}, 094522 (2004).
\bibitem{Kugler} M. Kugler, {\O}. Fischer, C. Renner, S. Ono, Y. Ando:
Phys. Rev. Lett. {\bf 86}, 4911 (2001).





















































%
%




\end{thebibliography}
%
%
%

%
%



\printindex
\end{document}